\documentclass[twocolumn]{aastex6}
\def\mkm{\mu{\rm m}}
\usepackage{ulem}
\begin{document}

\title{Mid-infrared extinction and fresh silicate dust
 towards the Galactic Center}

\author{Nikolai V. Voshchinnikov,\altaffilmark{1,2}
Thomas Henning\altaffilmark{3}, and Vladimir B. Il'in\altaffilmark{1,4,5}}
%
%

\altaffiltext{1}{Sobolev Astronomical Institute,
St.~Petersburg University, Universitetskii prosp. 28,
           St.~Petersburg, 198504, Russia}
\altaffiltext{2}{n$_{.}$voshchinnikov@spbu.ru}
\altaffiltext{3}{Max-Planck-Institut f\"ur Astronomie,
K\"onigstuhl 17, D-69117, Heidelberg, Germany}
\altaffiltext{4}{Main (Pulkovo) Astronomical Observatory,
 Pulkovskoe sh. 65, St. Petersburg, 196140, Russia}
\altaffiltext{5}{St.~Petersburg State University of Aerospace Instrumentation,
Bol. Morskaya 67, St. Petersburg, 190000, Russia}

\begin{abstract}
We interpret the interstellar extinction observed towards
the Galactic Center (GC) in the wavelength range $\lambda = 1 - 20\,\mkm$.
 Its main feature is the flat extinction at $3 - 8\,\mkm$ 
whose explanation is still a problem for the cosmic dust models.
We search for structure and chemical composition of dust grains
that could explain the observed extinction. In contrast
to earlier works we use laboratory measured optical constants
and consider particles of different structure.
 We show that a mixture of compact grains of aromatic carbon and of
 some silicate is better suited for reproducing the flat extinction
 in comparison with essentially porous grains or aliphatic carbon particles.
 Metallic iron should be located inside the particle, i.e. cannot
form layers on silicate grains as the extinction curves become then very
peculiar. 
 We find a model including aromatic carbonaceous
particles and three-layered particles with an olivine-type silicate
core, a thin very porous layer and a thin envelope of magnetite
that provides a good (but still not perfect) fit to the observational data.
 We suggest that such silicate dust should be fresh, i.e. recently
formed in the atmospheres of late-type stars in the
central region of the Galaxy.
We assume that this region has a radius of about 1 kpc
and produces about a half of the observed extinction.
 The remaining part of extinction is caused by a ``foreground'' material
 being practically transparent at $\lambda = 4 - 8\,\mkm$.
\end{abstract}

\section{Introduction} \label{intro}

The center parts of the Milky Way are a unique place to study
different processes in the vicinity of a supermassive black hole as well as
dynamics and star formation under extreme conditions
\citep{gen10,mg15}.
The Galactic Center (GC)\footnote{Hereafter, by the Galactic
 Center  we mean a central region of about 1 kpc radius.}
 invisible at the optical wavelengths can be observed in the infrared
 (IR) where the extinction amounts to $A_{\rm Ks}=2\fm49$ \citep{fritz11}.
 A distinguishing feature of the GC extinction
is its flat wavelength  dependence at  $3\,\mkm < \lambda < 8\,\mkm$.
 The flat or gray extinction in the GC was firstly measured by
\citet{luetal96} with {\it ISO}, using hydrogen recombination lines,
and confirmed by \citet{lutz99}, \citet{nishi09}, and \citet{fritz11}.
 Numerous recent observations appear to suggest the universality
of flat extinction in the mid-IR for both diffuse and dense environments
\citep[see][~for a summary]{wang13}.

\citet{fritz11} have compared different dust models capable of
explaining the mid-IR extinction in the GC.
 The models were from \citet{wd01} (mixture of carbonaceous and silicate
 spheres),
\citet{zu04} (mixture of carbonaceous and silicate particles and additionally
composite grains consisting of silicates, organic refractory material, water
ice, and voids\footnote{The optical properties of such particles were
calculated using the Mie theory for homogeneous spheres and refractive
indexes averaged according to the Effective Medium Theory (EMT).}),
\citet{dwek04} (mixture of bare particles of \citet{zu04} 
and additionally metallic needles), and
\citet{vihd06} (multi-layered spheres consisting of silicate, carbon, and
vacuum).   \citet{wang14} have developed the idea of \citet{dwek04} and
considered additionally micrometer-sized
particles from amorphous carbon, graphite, silicate or iron. 
 Such a model with amorphous carbon 
explained the flat extinction at $3- 8\,\mkm$, but required
 the solid-phase C abundance C/H$=$352~ppm that
exceeded the solar abundance of carbon \citep[269 ppm,][]{agss09}.
 An important feature of the modelings mentioned above is a priori selection
of the optical constants of grain materials. Moreover, all the
authors used the optical constants of the ``astronomical silicate''
(astrosil) obtained by empirical fits to some observations by \citet{dl84}.
 The imaginary part of the complex refractive index of astrosil $k$ slightly
grows with $\lambda$ in the region $3 - 8\,\mkm$, which does not coincide with the behaviour
of $k$ for 
any silicate material \citep[see Fig.~A.1 in][]{jones13}.

It should be emphasized that there are no cosmic dust models that can explain
the flat (excess) mid-IR extinction observed in the GC and other galactic objects. 
The COMP-AC-S model of \citet{zu04} gives a good fit,
but produces strong 3 $\mkm$ band, which disagrees with
the trend found in the Coalsack nebula Globule~7 by \citet{wang13}.
The most recent model of \citet{wang15} includes
$4\,\mkm$ clean water ice particles
and does explain both mid-IR extinction and the abundance of oxygen in dust,
but the ice particles hardly can be so large and clean in the
interstellar medium (ISM).

In this paper, 
we analyze a large set of dust models, concentrating on variations of
grain structure and a proper presentation of grains' chemical composition,
to find a model that fits the near- and mid-IR extinctions and
the 10~$\mkm$ feature observed to the GC.
 The next section contains a description of the observational data
and the models. The results and their discussion are presented in
Sect.~\ref{res}.
Concluding remarks are given in Sect.~\ref{concl}.

\section{Observational data and dust model} \label{obs}

The GC extinction has been observationally obtained
by \citet{fritz11} (the region 1.3--19 $\mkm$),
\citet{nishi09} ({1.2--8} $\mkm$), and 
\citet{chiti06} ({1.2--25} $\mkm$).
 The latter paper  contains the probable extinction profile of the 9.7~$\mkm$
silicate feature for the GC. All data have been normalized by us
in order to have $A_{\rm Ks}=2\fm49$ at $\lambda_{\rm Ks}=2.17\,\mkm$,
\begin{equation}
A^*(\lambda) = \frac{A(\lambda)}{A_{\rm Ks}} \, 2\fm49\,.
\label{eq1}
\end{equation}
They are plotted in all Figures below.

It should be note that the GC extinction was estimated from observations
in different ways. 
 As a result, the data of \citet{fritz11} were mainly derived for
the central 14$^{\prime\prime}\times$20$^{\prime\prime}$ region, 
the data of \citet{nishi09} are averaged over the region
$|l| < 3^\circ, |b| < 1^\circ$, 
and the data of \citet{chiti06} are the extinction towards
the Wolf-Rayet star WR 98a ($l \approx 358^\circ, b \approx 0^\circ$)
extended to the line of sight to GCS3.
However, the extinction law for $\lambda < 14\, \mkm$ is
practically the same.
 Hence, the data can be combined, and the question on where is located
 the dust that produces the extinction is not as important as it could be.

We base our analysis on the model of \citet{hv14}
who chose the initial size distributions of silicate and carbonaceous
dust that fited the mean Milky Way extinction curve with
$R_V=3.1$ \citep{wd01} and considered dust grain size evolution due
to the accretion and coagulation processes.
 
So, our model contains two populations of grains:
silicate (Si) and carbonaceous (C) ones\footnote{To reproduce the 2175\AA\,
feature small graphite spheres were also involved
\citep[see, e.g.,][]{dvi10}.}
with the size distributions from \citet{hv14}.
As extinction only weekly depends on
the particle shape \citep{vd08}, we assume that dust grains are spherical.

Thus, the model has the following parameters:
1) the chemical composition of silicate and carbonaceous particles;
2) the structure of particles;
3) the relative number of silicate grains
${\cal K}_{\rm Si}=N_{\rm Si}/N_{\rm dust}$,
where $N_{\rm Si}$ and $N_{\rm dust}$ are the column densities
of silicate grains and all dust particles, respectively;
4) the time of evolution.
Sometimes, we also included an additional population of dust.

When considering the chemical composition,
we mainly oriented on the optical constants obtained
in Jena laboratory (http://www.astro-uni-jena.de/Laboratory/).
 Information about these and many other data
is collected in the Heidelberg--Jena--Petersburg Database of
Optical Constants (HJPDOC) described by \citet{heal99} and \citet{jager03b}.
 The materials used for our modelling are outlined in
Table~\ref{mat} in the Appendix. 

For homogeneous spheres, the extinction efficiency factors were calculated 
with the Mie theory.
For composite particles, the factors were computed by using the Mie theory
and the Bruggeman mixing rule of the EMT or the theory for multi-layered
spheres \citep[see][]{vm99}.

\section{Results and discussion} \label{res}

\setcounter{table}{1}
\begin{table*}
\caption{Best-fit dust models\label{mod}}
\begin{tabular}{rlcccccl}
\tablewidth{0pt}
\hline
\hline
$N$ &\multicolumn{1}{c}{Model components\mbox{\hspace{4.5cm}}} & $\chi^2$/d.o.f. & $A^*(7.5\,\mkm)$ & {$R_V$} & $\lambda_m$ & $A^*(\lambda_m)$ & Figs. \\
\hline
\noalign{\smallskip}
 & Observations & -- & 0.81 & $\la 3$: & 9.6  & 3.45 & Figs.~\ref{f1}--\ref{f4}\\
\multicolumn{8}{c}{Homogeneous spheres}\\
\noalign{\smallskip}
1 & astrosil (${\cal K}_{\rm Si}=0.50$) / ACBE$\_$zu &   36.4 & 0.561 & 3.07 & 9.5  & 3.43 & Fig.~\ref{f1}\\
2 & olmg50 (${\cal K}_{\rm Si}=0.25$) / cell400      &  116.4 & 0.320 & 3.25 & 9.8  & 3.41 & Fig.~\ref{f1}\\
3 & olmg50 (${\cal K}_{\rm Si}=0.55$) / cell1000     &   36.6 & 0.321 & 3.26 & 9.8  & 3.36 & Fig.~\ref{f1}\\
4 & olmg40 (${\cal K}_{\rm Si}=0.77$) / cell1000     &  112.9 & 0.261 & 3.29 & 9.8  & 3.46 &\\
5 & olmg100 (${\cal K}_{\rm Si}=0.56$) / cell1000    &   36.1 & 0.706 & 2.61 & 9.7  & 3.46 &\\
6 & pyrmg50 (${\cal K}_{\rm Si}=0.47$) / cell1000    &   47.8 & 0.381 & 2.87 & 9.2  & 3.47 & Fig.~\ref{f1}\\
7 & pyrmg40 (${\cal K}_{\rm Si}=0.48$) / cell1000    &   43.1 & 0.386 & 3.00 & 9.0  & 3.49 &\\
8 & pyrmg100 (${\cal K}_{\rm Si}=0.39$) / cell1000   &   74.9 & 0.385 & 2.73 & 9.4  & 3.48 &\\
9 & OHM-SiO (${\cal K}_{\rm Si}=0.92$) / cell1000    &   35.4 & 0.503 & 4.02 &10.0  & 3.41 &\\
10 & olmg50 (${\cal K}_{\rm Si}=0.44$) / H$_2$O (${\cal K}_{\rm H_2O}=0.20$) / cell1000  & 30.4 & 0.420 & 2.98 & 9.8  & 3.42 & \\
11 & olmg50 (${\cal K}_{\rm Si}=0.50$) / Fe (${\cal K}_{\rm Fe}=0.20$) / cell1000  & 87.8 & 0.208 & 3.94 & 9.8  & 1.84       & \\
\noalign{\smallskip}
\multicolumn{8}{c}{EMT-Mie calculations}\\
\noalign{\smallskip}
12 & 80\%olmg50+20\%\,vac (${\cal K}_{\rm Si}=0.53$) / cell1000     &  32.9  & 0.338 & 2.64 &10.0  & 2.86           &\\
13 & olmg50 (${\cal K}_{\rm Si}=0.46$) / 80\%cell1000+20\%\,vac     &  26.4  & 0.383 & 3.34 & 9.8  & 2.78           &\\
14 & a-Sil$_{\rm Fe}$ (${\cal K}_{\rm Si}=0.40$) / cell1000         &  37.2  & 0.312 & 4.60 & 9.9  & 1.40           &\\
15 & a-Sil$_{\rm Fe}$ (${\cal K}_{\rm Si}=0.34$) / optEC$_{\rm (s)}$& 487.0  & 0.080 & 3.92 &10.0  & 3.41 &\\
16 & amFo-10Fe30FeS (${\cal K}_{\rm Si}=0.47$) / cell1000           &  40.6  & 0.287 & 4.29 & 9.9  & 1.64             &\\
17 & amEn-10Fe30FeS (${\cal K}_{\rm Si}=0.39$) / cell1000           &  36.5  & 0.290 & 5.06 & 9.5  & 1.77             &\\
\noalign{\smallskip}
\multicolumn{8}{c}{Core-mantle spheres}\\
\noalign{\smallskip}
18 & 20\%\,vac--80\%\,olmg50 (${\cal K}_{\rm Si}=0.53$) / cell1000     & 32.6 & 0.335 & 2.66 &10.0  & 2.82 &\\
19 & olmg50 (${\cal K}_{\rm Si}=0.44$) / 20\%\,vac--80\%cell1000       & 25.9 & 0.380 & 3.68 & 9.8  & 2.53 &\\
20 & 93\%\,a-Sil$_{\rm Fe}$--7\%\,cel1000 (${\cal K}=0.28$) / 73\%\,optEC$_{\rm (s)}$--27\%\,cell1000   & 313.9 & 0.240 & 3.48 & 10.0  & 3.44 & Fig.~\ref{f4}\\
21 & 93\%\,olmg50--7\%\,cel1000 (${\cal K}=0.19$) / 73\%\,cel400--27\%\,cell1000   & 90.1 & 0.391 & 2.99 & 9.8  & 3.35 & \\
\noalign{\smallskip}
\multicolumn{8}{c}{Three-layered spheres}\\
\noalign{\smallskip}
22 & 10\%\,Fe--10\%\,vac$^*$--80\%\,olmg50 (${\cal K}=0.77$) / cell1000          & 164.3  &  0.231  & 3.06   &  9.8   &  3.41  & Fig.~\ref{f2}\\
23 & 10\%\,vac--10\%\,Fe--80\%\,olmg50 (${\cal K}=0.18$) / cell1000              & 68.5  &  0.280    & 5.29  &  9.6    & 0.68   & Fig.~\ref{f2}\\
24 & 98.99\%\,olmg50--1\%\,vac$^*$--0.01\%\,Fe (${\cal K}=0.09$) / cell1000      & 89.5  & 1.213  & 4.83  &  8.2 &  1.49 & Fig.~\ref{f2}\\
25 & 90\%\,olmg50--5\%\,vac$^*$--5\%\,Fe$_3$O$_4$ (${\cal K}=0.91$) / cell1000   & 6.7  & 0.696 & 3.22 & 9.8 & 3.45 & Fig.~\ref{f3}, \ref{f4}\\
26 & 90\%\,olmg50--5\%\,vac$^*$--5\%\,Fe$_2$O$_3$ (${\cal K}=0.47$) / cell1000   & 31.7  & 0.332 & 3.56 & 9.8 & 2.39 & Fig.~\ref{f3}\\
27 & 90\%\,olmg50--5\%\,vac$^*$--5\%\,FeO (${\cal K}=0.48$) / cell1000           & 36.2  & 0.319 & 3.51 & 9.8 & 2.21 & Fig.~\ref{f3}\\
28 & 90\%\,olmg50--5\%\,vac$^*$--5\%\,FeS (${\cal K}=0.16$) / cell1000           & 94.3  & 0.319 & 4.42 & 9.8 & 0.61 & Fig.~\ref{f3}\\
29 & 90\%\,pyrmg50--5\%\,vac$^*$--5\%\,Fe$_3$O$_4$ (${\cal K}=0.72$) / cell1000  & 75.7  & 0.781 & 2.91 & 9.1 & 3.22 & \\
\noalign{\smallskip}
\multicolumn{8}{c}{Two-cloud model}\\
\noalign{\smallskip}
30 & model 20 + model 25 (see Sect.~\ref{fore})   & 84.9  & 0.468 & 3.32 & 9.9 & 3.41 & Fig.~\ref{f4}\\
\noalign{\smallskip}
\hline
\noalign{\smallskip}
\end{tabular}
NOTES.
Column 3: fit goodness $\chi^2$/d.o.f.,
where d.o.f. means the degree of freedom (we took it equal to 24);
Column 4: normalized extinction at $\lambda=7.5\,\mkm$;
Column 5: ratio of the total-to-selective extinction;
Column 6: position of the 10 $\mkm$ peak in $\mkm$;
Column 7: normalized strength of the 10 $\mkm$ peak;
vac$^*$ --- very porous layer.
\end{table*}

We varied the model  parameters
to fit the observed GC extinction.
 The number of possible model variants is very large, 
but it can be significantly
reduced by applying available knowledge on the physics 
of dust formation, growth and evolution
\citep[see, e.g.,][]{chiar13, jones13, gs14}.

Information about some models considered
is collected in Table~\ref{mod} which
gives a description of the model (column 2),
normalized $\chi^2$ characterizing the goodness
of the fit for 29 observational points
from \citet{fritz11} and \citet{nishi09} (column 3),
obtained values of normalized extinction at
$\lambda=7.5\,\mkm$ $A^*(7.5\,\mkm)$
(assuming $A_{\rm Ks}=2\fm49$, column 4),
$R_V$ (ratio of the total-to-selective extinction, column 5)
and the position and strength of the 10 $\mkm$ peak (columns 6 and 7).

The fitting procedure was as follows.
First, we fitted the extinction shortward 8.8 $\mkm$, i.e.
19 points from \citet{fritz11} and all 10 points from \citet{nishi09}. 
 The values of the normalized $\chi^2$ given in Table~\ref{mod}
just characterize this fitting.
 Then, by varying the fraction
of silicate grains ${\cal K}_{\rm Si}$, we fitted the relative strength
of the 10~$\mkm$ band. 
 The position of the band was mainly fitted
by the proper choice of the silicate material.
 The relative strength
and position of the band at 10 $\mkm$ were taken from \citet{chiti06}.
 Note that their data for the 18 $\mkm$ band are less reliable
and that many silicates have the bending bands in the range
$16 - 23\,\mkm$ \citep[see][]{he10}. Therefore,
we did not model the 18\,$\mkm$ band.

\subsection{Homogeneous particles} \label{hom_part}

\begin{figure}
\centerline{
\resizebox{8.5cm}{!}{\includegraphics{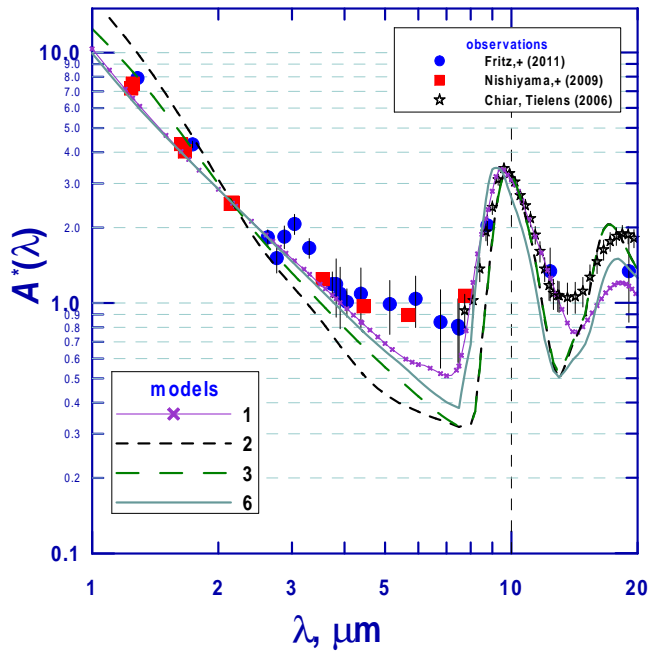}}
}
\caption{Comparison of the normalized IR extinction observed towards
the Galactic Center with that predicted by the models with homogeneous
particles. The model numbers are from Table~\ref{mod}.}
\label{f1}
\end{figure}

We have considered a number of two- and three-component models with compact
homogeneous grains. 
 We started with the standard mixture of grains of artificial silicate,
astrosil, and amorphous carbon
ACBE$\_$zu (model 1, see Table~\ref{mod} and Fig.~\ref{f1}).
As expected,
the wavelength dependence of extinction derived was steeper
than that given by observations.

The next step was to find better models by variations of the
laboratory optical constants.
 A comparison between the models with the aliphatic and aromatic carbon  
(models 2, 3) showed that the near-IR ($\lambda < 2\,\mkm$) and mid-IR
extinction was much better reproduced by the model with aromatic
carbon cell1000.
So, we chose this material as the basic one in the subsequent modelling.

 Note that the carbon materials cell400 and cell1000 used by us differ
in the degree of ``graphitization'' \citep{jetal98}. 
 Therefore, for the former, in first approximation the imaginary part of the
refractive index  $k \sim \lambda^{-1}$ for $\lambda = 1-10\, \mkm$, while
for the latter, $k \approx {\rm const.}$ (for graphite,
$n, k$ grow with $\lambda$).
 Obviously, such graphitization favours excess IR extinction.

Further, we examined different types of silicates: olivines and pyroxenes
with different content of Mg and Fe (models 3 -- 8).
 As can be seen, olivines better explain the observations as the
silicate peak produced by pyroxenes is shifted to $\lambda =9.0 - 9.4\,\mkm$
(Table~\ref{mod} and Fig.~\ref{f1}).
 Though forsterite grains (model 5) well fit the observed mid-IR extinction,
in this case dust grains contain no iron, which is hardly probable
according to contemporary understanding of cosmic dust origin and evolution.
 Our attempts to add iron or water ice as the third component into
our silicate-carbonaceous mixture (models 10, 11) failed as mid-IR extinction
always became steeper.

\subsection{EMT-Mie calculations and core-mantle particles}\label{emt_cm}

The physical conditions in which dust grains originate and grow
should lead to formation of heterogeneous particles, in particular, porous.
Two grain structures are generally expected:
layered particles corresponding to subsequent accretion of different species,
and an alternative --- particles with small more or less randomly
distributed inclusions.
 In the former case the optical properties of heterogeneous particles
 are modelled with
 the Mie-like theory for layered particles (in particular, core-mantle),
in the latter case by using homogeneous particles with the averaged
dielectric functions (EMT-Mie calculations).
 
 We present four models with porous\footnote{The volume factions
of vacuum and a solid material are 20\% and 80\%, respectively.}
silicate or carbonaceous particles (models 12, 13 and 18, 19) 
to illustrate that the porosity does not make the fitting much better
in a comparison with compact grains, 
but leads to the shift and decrease of the silicate peak.

We have also considered the models with the refractive indexes constructed by
\citet{jones12}, \citet{jones13} and \citet{kohler14} (models 14 -- 17).
 None of these models produces the flat mid-IR extinction with the worst
fit to the data given by aliphatic carbon optEC$_{\rm (s)}$ (model 15).
 The models 20 and 21  with core-mantle grains give a good opportunity to
test the hypotheses of \citet{jones13} who predicted that in the diffuse
ISM large a-C:H grains are to be covered by a a-C 20 nm thick envelope
and large Si grains by a a-C 5 nm thick envelope. As can be seen, in
this case extinction is inconsistent with the observations of the flat
mid-IR extinction in the GC.

Note that an increase of the thickness of the a-C mantles of
silicate grains leads to an increase of the mid-IR extinction as has
been demonstrated in the work of \citet{kohler14},
where they have also analyzed the effect of ``dirtiness'' of silicates
(due to absorbing inclusions of FeS). However, this increase is
certainly not enough, while the thick a-C mantles begin to affect
the silicate bands strength (see their Fig.~3).

\subsection{Three-layered particles} \label{3_part}

\begin{figure}
\centerline{
\resizebox{8.5cm}{!}{\includegraphics{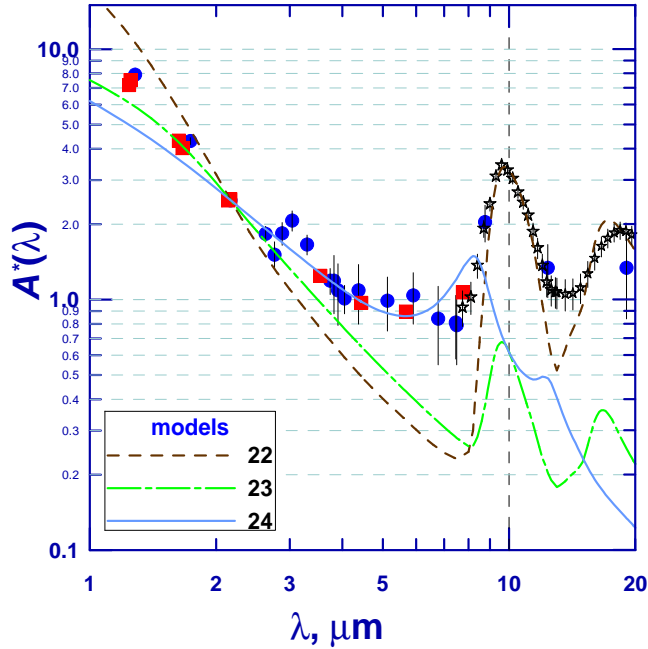}}
}
\caption{Same as Fig.~\ref{f1} but for the models with mixtures of
the three-layered silicate particles and homogeneous carbonaceous
particles. Three-layered particles consist of olivine, iron and vacuum.
Fe is located in the particle core (model 22), intermediate layer
(model 23) or outer layer (model 24).
}
\label{f2}
\end{figure}

The use of multi-layered particles permits a more
sophisticated treatment of the processes of grain growth and evolution.
Specifically, it is possible to analyse the role of iron which is
one of the major dust-forming elements \citep{jones00,dwek16}.
 The abundance of iron in the solid-phase of the ISM may reach
 97 -- 99\% of the cosmic abundance \citep{vh10}.
 Iron can be incorporated into dust grains in the form of oxides
(FeO, Fe$_2$O$_3$, Fe$_3$O$_4$),
(Mg/Fe)-silicates, sulfide (FeS), and metallic iron.
 The last two cases come from the contemporary theory of dust condensation
in  circumstellar environments.
 \citet{gs99,gs14} note that Fe and FeS start to condense
at temperatures well below the stability limits of silicates
like forsterite and enstatite. 
 This should lead to formation of layered particle.
 At low temperatures, the conversion of solid
iron into iron oxides may occur \citep[][ p.~306]{gs14}.

Figure~\ref{f2} shows the wavelength dependence of extinction for
the models with three-layered particles including of olivine and vacuum.
 Iron is located in the particle core (model 22), intermediate layer
(model 23), or outer layer (model 24).
 As seen, the presence of metallic iron at any place
inside a particle, excluding its core, drastically changes
extinction --- iron totally screens the underlying
layers and influences the optics of the overlying ones.
As a result, one cannot properly reproduce
either the position and shape of the observed silicate band (models 23, 24)
or the slope of the wavelength dependence of IR extinction (models 22, 23).

\begin{figure}
\centerline{
\resizebox{8.5cm}{!}{\includegraphics{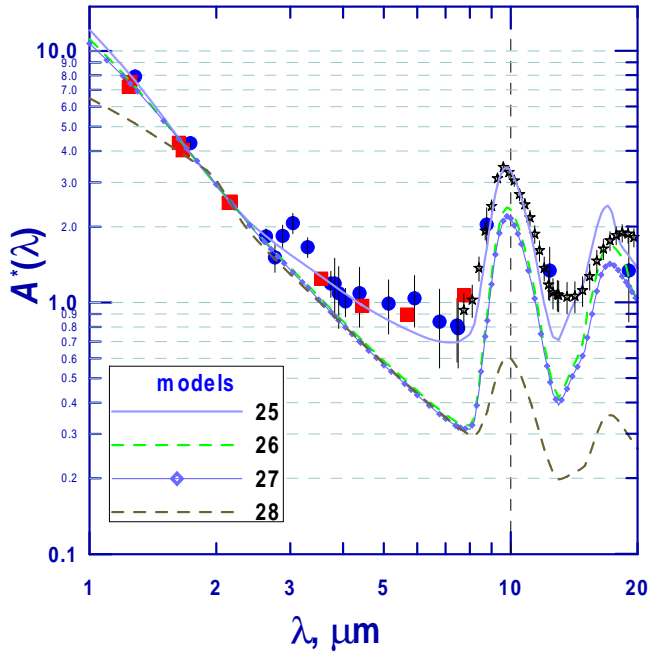}}
}
\caption{Same as Fig.~\ref{f2} but for three-layered particles
consisting of olivine (core), very porous intermediate layer
and iron oxide or sulfide (outer layer). The outer layer is from
Fe$_3$O$_4$ (model 25), Fe$_2$O$_3$ (model 26), FeO (model 27)
or FeS (model 28).
}
\label{f3}
\end{figure}

However, iron can be oxidized or sulfidized, 
which opens a way to explain the observations.
 Figure~\ref{f3} shows
the extinction calculated for four models with olivine particles (olmg50)
coated by a thin very porous layer
and a thin (2 -- 3 nm thick) envelope of iron oxide or iron sulfide.
 It is evident that the model 25 with
magnetite agrees closely with the observational data.
 This model well reproduces near-IR extinction and the 10\,$\mkm$
 peak and gives nearly as large mid-IR extinction as observed.
The model also produces the visual extinction $A_V/A_{Ks}=15.2$
which is close to the observed median value equal to $13.4$ \citep{natall16}.

 Note that the replacement of olivine with pyroxene (model 29)
leads to even a better coincidence with the
observed extinction at $\lambda =5 - 9\,\mkm$ but does not allow one 
to explain properly the near-IR extinction and the position of the 
silicate feature.

So, we see that the model 25 is practically the only way of
successful fitting of the data, when keeping in mind available
information on cosmic dust.
  Considering the model 25 with the particles from olivine olmg50 and
amorphous carbon cell1000 as a {\it prototype} of possible dust models
for the GC.

\subsection{Foreground extinction} \label{fore}

In previous modelling we ignored the distribution of the extinction
along the line of sight. However, the 3-dimensional extinction map of
the GC shows that about half of the extinction in the sightlines
of \citet{nishi09} is reached in a distance of about 5 kpc from the Sun
\citep[see Fig.~9 in][]{schu14}.

\begin{figure}
\centerline{
\resizebox{8.5cm}{!}{\includegraphics{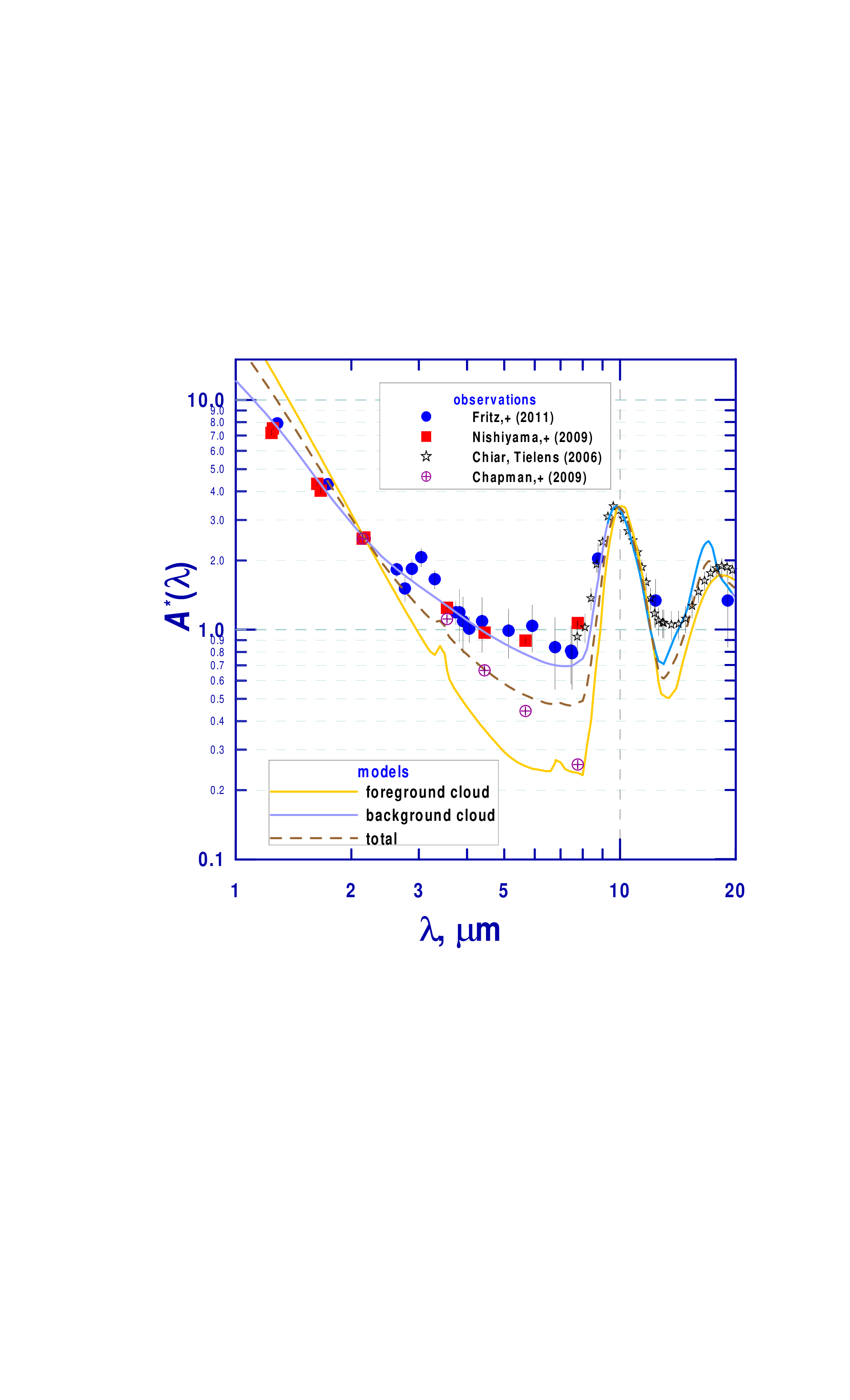}}
}
\caption{Comparison of the normalized IR extinction observed 
towards the Galactic Center (filled circles and squares)
with that predicted by the model 20 with processed core-mantle
particles in the foreground cloud and model 25 with fresh silicate
3-layered particles in the background cloud. Dashed brown line
shows the total extinction produced in the foreground and background
clouds as calculated from Equation~(\ref{eq2}).
Crossed circles present the average wavelength dependence of the
interstellar extinction for three molecular clouds in the local ISM
\citep[][]{chap09}.}
\label{f4}
\end{figure}

We assume different populations of dust grains in the foreground
dusty complexes and in the central galactic region\footnote{Note
that a model based on combination of three regions with different
extinction curves has been considered by \citet{gao13}.}
 Most likely the dust producing the foreground extinction is processed,
in particular, the silicate grains are covered by carbon \citep[][]{jones13}.
 Such grains are properly described by the theoretical model 20
and give very low relative extinction in the mid-IR
\citep[$A(7.5\,\mkm)/A_{\rm Ks} \sim 0.1 - 0.2$,
see Fig.~\ref{f4} and][]{kohler14}.
However, there exist several places in the Galaxy
where low mid-IR extinction has been observed. On Fig.~\ref{f4} we plotted
the average extinction for three molecular clouds obtained by
\citet{chap09}. It is visible that the model 20 agrees roughly with the
measurements.

For the central galactic region, we applied the model 25
with freshly formed silicate dust.
The total extinction for our ``two-cloud'' model was calculated as
\begin{equation}
A_{\rm total}^*(\lambda) =
f\,A^*_{\rm foregr}(\lambda) + (1-f)\,A^*_{\rm backgr}(\lambda),
\label{eq2}
\end{equation}
where $f$ is the contribution of the foreground clouds to the total extinction.
At the moment, the available data \citep[see, e.g.,][]{schu14}
do not allow one to estimate the value of $f$ with a sufficient accuracy,
therefore, we just use 0.5 for simplicity.

Figure~\ref{f4} and Table~\ref{mod} show
the extinction produced by the two-cloud model
($f=0.5$). Its agreement with the observational data is not perfect
but good enough.

\section{Concluding remarks} \label{concl}

According to the modern ideas on cosmic dust evolution in
the diffuse ISM,  the silicate grains should be covered by a significant
envelope from amorphous carbon on a short time scale \citep{jones13}.
 Moreover, amorphous olivine MgFeSiO$_4$ (olmg50) is a possible mineral 
in dust grains forming in the atmospheres of late-type giants, but 
it is not believed to be the main material of silicate 
particles in the ISM \citep[see, e.g.,][]{jager03b}.
Therefore, we suggest that silicate dust in our model is ``fresh'',
i.e. recently formed in the atmospheres of the late-type stars in the GC.
Our suggestion is rather natural as the GC is dominated by old stars.


Carbonaceous particles are more processed in comparison with
silicate ones that is determined by lower efficiency of their destruction
\citep[][]{sla15}. Intense radiation fields in the GC are favourable
for the fast photo--dissociative aromatisation of a-C(:H) materials
\citep[][]{jones13, jones14}.

Obviously, the model found by us does not fit the data perfectly and
one cannot exclude other possible solutions to the
problem of the flat mid-IR extinction towards the GC.
However, we pay attention to the potential of our approach ---
to relate the problem solution with specific structure and 
composition of dust grains relying the
laboratory data on optical constants and
contemporary ideas on cosmic dust grain evolution.

\acknowledgments
We are grateful to anonymous referee for very useful remarks.
We thank A. P. Jones and M. K\"ohler for sending us
the refractive indexes in the tabular form.
We are grateful to Harald Mutschke for numerious consultations.
NVV and VBI acknowledge the support from RFBR grant 16-02-00194
and RFBR--DST grant 16-52-45005.



\appendix
\begin{table*}[hb]
\centering
\caption{Sources of optical constants}\label{mat}
\begin{tabular}{lll}
\tablewidth{0pt}
\hline
\hline
{Notation} & {Material} & {Reference}\\
\hline
astrosil & astronomical silicate & \citet{d03} \\
olmg50 & amorphous olivine (MgFeSiO$_4$)& \citet{dor95} \\
olmg40 & amorphous olivine (Mg$_{0.8}$Fe$_{1.2}$SiO$_4$) & \citet{dor95} \\
olmg100& amorphous olivine (Mg$_{2}$SiO$_4$, forsterite) & \citet{jager03a} \\
pyrmg50 & amorphous pyroxene (Mg$_{0.5}$Fe$_{0.5}$SiO$_3$)& \citet{dor95} \\
pyrmg40 & amorphous pyroxene (Mg$_{0.4}$Fe$_{0.6}$SiO$_3$) & \citet{dor95} \\
pyrmg100 & amorphous pyroxene  (MgSiO$_3$, enstatite)  & \citet{dor95} \\
OHM-SiO  & O-rich interstellar silicate & \citet{ohm92} \\
a-Sil$_{\rm Fe}$ & amorphous olivine (MgFeSiO$_4$+10\%Fe) & \citet{jones13} \\
amFo-10Fe30FeS & amorphous forsterite (Mg$_2$SiO$_4$+10\%Fe+30\%FeS)& \citet{kohler14} \\
amEn-10Fe30FeS & amorphous enstatite (MgSiO$_3$+10\%Fe+30\%FeS)& \citet{kohler14} \\
\noalign{\smallskip}
ACBE$\_$zu  & amorphous carbon (type BE)& \citet{zu96} \\
cell400 & pyrolizing cellulose ($T=400\degr$C, aliphatic, a-C(:H)) & \citet{jetal98} \\
cell1000 & pyrolizing cellulose ($T=1000\degr$C, aromatic, a-C) & \citet{jetal98} \\
optEC$_{\rm (s)}$ & amorphous carbon (a-C(:H), band gap $E_g=2.5$~eV)& \citet{jones12} \\
\noalign{\smallskip}
Fe & iron & \citet{jones13} \\
FeO & w\"ustite & \citet{he95} \\
Fe$_2$O$_3$ & hematite & Jena laboratory \\
Fe$_3$O$_4$ & magnetite & Jena laboratory \\
FeS & troilite & \citet{pol94} \\
\noalign{\smallskip}
H$_2$O & water ice & \citet{warren08}\\
\hline
\end{tabular}
\end{table*}

\end{document}